\documentclass[sigconf]{acmart}
\usepackage{multirow}
\usepackage{graphicx}
\usepackage{enumitem}
\usepackage{subfigure}  
\usepackage{bbding} 
\usepackage{stfloats}
\usepackage{float}
\usepackage{makecell}
\usepackage{algorithm} 
\usepackage{algpseudocode}
\usepackage{bbding}
\usepackage{balance}

\begin{document}
\title{Time Matters: Enhancing Pre-trained News Recommendation Models with Robust User Dwell Time Injection} 

\author{Hao Jiang}
\affiliation{ %
  \institution{Communication University of China,}
  \city{Beijing}
  \country{China}}
  \email{jianghao11@cuc.edu.cn}

\author{Chuanzhen Li}
\affiliation{ %
  \institution{Communication University of China,}
  \city{Beijing}
  \country{China}}
  \email{lichuanzhen@cuc.edu.cn}

\author{Mingxiao An}
\affiliation{ %
  \institution{Microsoft,}
  \city{Redmond, Washington}
  \country{USA}}
  \email{mingxiaoan@microsoft.com}

\begin{abstract}
Large Language Models (LLMs) have revolutionized text comprehension, leading to State-of-the-Art (SOTA) news recommendation models that utilize LLMs for in-depth news understanding. Despite this, accurately modeling user preferences remains challenging due to the inherent uncertainty of click behaviors. Techniques like multi-head attention in Transformers seek to alleviate this by capturing interactions among clicks, yet they fall short in integrating explicit feedback signals. \textbf{User Dwell Time} emerges as a powerful indicator, offering the potential to enhance the weak signals emanating from clicks. Nonetheless, its real-world applicability is questionable, especially when dwell time data collection is subject to delays. To bridge this gap, this paper propose two novel and robust dwell time injection strategies, namely \underline{Dwe}ll time \underline{W}eight (\textbf{DweW}) and \underline{Dwe}ll time \underline{A}ware (\textbf{DweA}). \textbf{DweW} concentrates on refining  \emph{Effective User Clicks} through detailed analysis of dwell time, integrating  with initial behavioral inputs to construct a more robust user preference. \textbf{DweA} empowers the model with awareness of dwell time information, thereby facilitating autonomous adjustment of attention values in user modeling. This enhancement sharpens the model's ability to accurately identify user preferences. In our experiment using the real-world news dataset from MSN website, we validated that our two strategies significantly improve recommendation performance, favoring high-quality news. Crucially, our approaches exhibit robustness to user dwell time information, maintaining their ability to recommend high-quality content even in extreme cases where dwell time data is entirely missing.
\end{abstract}



\maketitle

\section{Introduction}
Online news platforms have now emerged as the primary source for daily news for a vast number of users. With the large amount of news available every day, it's crucial to have a recommendation system that understands what users prefer. Fortunately, numerous powerful news recommendation systems have emerged recently. They mainly have two important parts: \textbf{User Interest Modeling}~\cite{ge2020graph, hu2020graph2, hu2020graph, an2019neural} and \textbf{News Understanding Modeling}~\cite{wu2019npa, wu2019neural, wu2019}. These parts focus on accurately figuring out user preferences and understanding news content.

\begin{figure}[tp]
  \includegraphics[width= \linewidth]{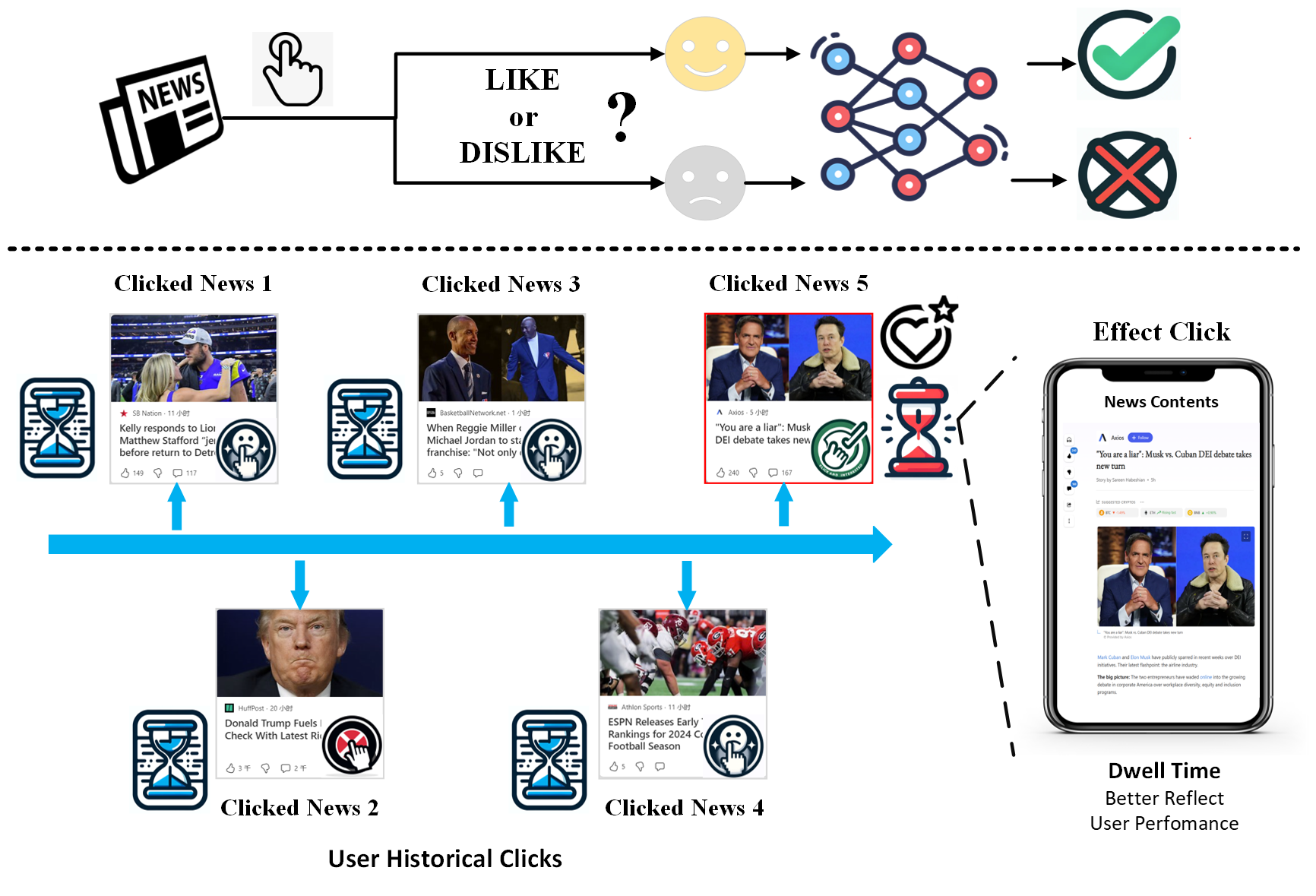}
  \caption{\textbf{Dwell time as a explicit feedback signal to refine user historical behaviors.}}
  \label{f1}
\end{figure}

Existing news recommendation methods mainly rely on user click behavior to construct user models. Unfortunately, user click behavior does not always accurately reflect their true interests. In the real world, the reasons for user clicks are varied and complex, as shown in bottom left of Fig. \ref{f1}. For example, users may quickly leave a news content page due to a "mis-click"~\cite{wang2021clicks}. Furthermore, user click behavior is not always a comprehensive representation of their interests~\cite{wang2022target}. For example, they might be attracted by the news cover or headline, but quickly realize after clicking that the content does not align with their interests and leave. Such noisy behaviors significantly impact the process of modeling user interests.

Fortunately, \textbf{User Dwell Time} can serve as an effective supplement to address the shortcomings in current user modeling methods. As demonstrated in Figure \ref{f1}, users spend significantly more time reading news about \emph{Elon Musk and Cuban's Defense} compared to other topics, such as \emph{Football Game} or \emph{Trump's News}. This disparity in dwell time suggests that Elon Musk-related news resonates more with users' genuine interests than other click behaviors. In summary, user dwell time can aid in filtering out irrelevant news and assist in accurately gauging the relevance of specific click behaviors in user modeling.

Several approaches highlight the importance of user dwell time in improving user modeling. These methods employ dwell time in two key areas: for evaluating news content quality and engagement on the news side~\cite{wu2022quality}, and for differentiating between positive and negative user interactions on the user side, enabling more refined modeling of user behavior~\cite{wu2020neural}. However, we identify three limitations in the existing dwell time injection methods. 
\begin{enumerate}[leftmargin=14pt,topsep=0pt]
  \item Relying solely on dwell time as an absolute measure of news quality and positive/negative user behavior fails to account for the varied and personal reading habits of individual users.
  \item Interpreting negative user behavior presents ambiguity, as short dwell times cannot be conclusively interpreted as disinterest or a lack of engagement with the content.
  \item An over-reliance on dwell time, without considering issues such as potential delays in data collection in real-world scenarios, can compromise the model's robustness. This neglects the complex and dynamic nature of user interactions in a live environment, where factors such as context, momentary distractions, and user moods play a pivotal role in shaping user engagement.
\end{enumerate}

To address this research gap, our paper presents a new dwell time-injection news recommendation method that leverages user dwell time to more precisely capture real user interests. We have developed two distinct mechanisms: \underline{Dwe}ll time \underline{W}eight (\textbf{DweW}) for pre-injection and \underline{Dwe}ll time \underline{A}ware (\textbf{DweA}) for post-injection of user dwell time, respectively. In \textbf{DweW}, we refine traditional approaches for capturing user behavior sequences by identifying "effective user click behaviors" through dwell time analysis. These effective behaviors, together with the original ones, are input into the \textbf{same} user encoder, producing two unique user representation vectors. We posit that effective behaviors are instrumental in sharpening the depiction of user interests. o further personalize to individual reading preferences, we introduced a reading preference network. A gating network merges these vectors to form a more nuanced user interest representation. In \textbf{DweA}, to allow the model to inherently understand dwell time and enable self-learning, we integrated dwell time data into a multi-head attention network from the outset. This technique combines temporal modeling with semantic analysis, enabling the model to weigh the semantic content of clicked news alongside the dwell time. This dual consideration allows the model to accurately identify "effective" user behaviors and precisely gauge user interests. Significantly, when applied to real-world MSN dataset, our models outperform baseline models on two datasets: \textbf{Normal Dataset} and \textbf{Real Dataset}. Additionally, it displayed exceptional robustness, maintaining high recommendation accuracy even in the absence of user dwell time data.

In summary, our major contributions are:
\begin{itemize}[leftmargin=0pt,topsep=0pt]
\item We systematically analyzed the distribution of user dwell time on real-world datasets and showcased its importance for precise user modeling.
\item We develop two innovative strategies for dwell time infusion aimed at boosting news recommendation accuracy and maintaining model robustness against fluctuations in user dwell time.
\item We demonstrate our model's effectiveness with tests on a real Feeds news dataset, where it showed marked enhancements in recommendation quality and maintained precise predictions despite variations in user dwell time.
\end{itemize}

\section{Related work}

Our works is related to news recommendation and dwell time-injection recommendation.

\subsection{News Recommendation}
In the realm of news recommendation systems, a variety of methodologies have been explored, each evolving to address the limitations of its predecessors. Feature-based methods \cite{Das2007,resnick1994grouplens,gershman2011news,goossen2011news,capelle2012semantics,lommatzsch2014real,claypool1999combining,saranya2012personalized,liu2010personalized}, which include collaborative filtering and content-based approaches, are foundational but struggle with issues like data sparsity and a heavy reliance on the quality of news content. This paved the way for deep learning-based methods\cite{wu2019npa,wu2019neural,wu2019,an2019neural,castelvecchi2016can}, which utilize neural network to enhance news understanding. However, these approaches grapple with challenges such as insufficient representation learning.Responding to these challenges, knowledge-enhanced approaches \cite{wang2017knowledge, bordes2013translating, ji2015knowledge,lin2015learning, sun2011pathsim, yang2022ekpn, zhao2017meta, wang2018ripplenet, wang2019kgat,wang2019knowledge,wang2018dkn,zhu2019dan,qi2021personalized,liu2020kred} were developed, which leverage external knowledge to improve interpretability and data handling. Yet, graph-based methods\cite{ge2020graph,hu2020graph,hu2020graph2,santosh2020mvl,qi2021hierec,wu2021user} try to fully capitalize on the role of entity relationships, advance the understanding of user-news interactions. 

The aforementioned methods fail to account for the "effectiveness" of user behaviors, as misleading click activities may adversely affect the model's recommendation performance.

\subsection{Dwell Time Injection Recommendation}
As we know, dwell Time can be a strong signals to evaluate news quality and enhance user implicit feedback. Therefore, studies on dwell time injection in recommendation systems has evolved significantly, particularly in news recommendation. The foundational work by Yi et al. \cite{yi2014beyond} introduced the concept of accurately computing and normalizing dwell time in user modeling, influencing later studies. Within the news recommendation field, the study by Wu et al. \cite{wu2020neural}, focused on split user click behaviors into positive and negative behaviors through dwell time threshold. This approach was further developed in the study by Wu et al. \cite{cprs}, which using dwell time information to revise attention value in user modeling. Besides, in news side, Wu et al. \cite{wu2022quality} define news quality with dwell time and integrate news quality into user modeling. Additionally, Xie et al. \cite{xie2023reweighting} expanded the use of dwell time by defining a new behavior, "valid read", which selects high-quality click instances and proposes a normalized dwell time function to adjust click signals during training. These studies underscore the role of dwell time in understanding user behavior and improving recommendation accuracy. 

However, these methods may not always be practical, given that dwell time associated with clicks can experience delays in real-world recommendation systems. The failure to precisely capture dwell time can significantly impact model performance. This paper concentrates on improving the robustness of time-injection models and effectively incorporating dwell time to address these challenges.

\section{Preliminaries}
\subsection{Record dwell time in industry}
In this section, we describe the process of collecting user dwell time in industrial scenarios and the method for converting it into discrete time intervals for model training.

\subsubsection{Recording dwell time}
In MSN, dwell time refers to the cumulative duration a user spends engaging with content. 
The calculation of dwell time relies entirely on user client events. 
MSN content can be categorized into three types: \emph{articles}, \emph{galleries}, and \emph{videos}.
For articles and galleries, dwell time is gauged by a timer that halts when the user navigates away from the page. In contrast, the dwell time for videos corresponds to the total playback duration of the video.

\subsubsection{Usage of dwell time}
In real-world settings, the dwell time we record is typically continuous, making it challenging for models to extract sufficient information from such data. Our approach involves segmenting user dwell time into discrete intervals, creating a space measured in seconds, and then categorizing it. This method facilitates the model's ability to access dwell time information via the embedding layer and allows it to perceive delays in dwell time (\( t \) is unknown). The dwell times referenced throughout this paper have been discretized using Eq. \ref{eq1}, where \( t \) denotes the actual dwell time.
\begin{equation}
\label{eq1}
\text{discrete\_dwell}(t) = 
    \begin{cases} 
    1 & \text{if } t \text{ is unknown}, \\
    2 & \text{if } t = 0, \\
    3 & \text{if } t \in (0, 5), \\
    \left\lfloor \frac{t}{5} \right\rfloor + 3 & \text{if } t \in [5, 60), \\
    \left\lfloor \frac{t}{60} \right\rfloor + 5 & \text{if } t \in [60, 600), \\
    9 & \text{if } t \geq 600.
    \end{cases}
\end{equation}

\subsection{User Dwell time Analysis}

\begin{figure*}[t]
   \centering
    \subfigure[Dwell Time Ratio Line Graph]{
        \includegraphics[width=0.23\linewidth]{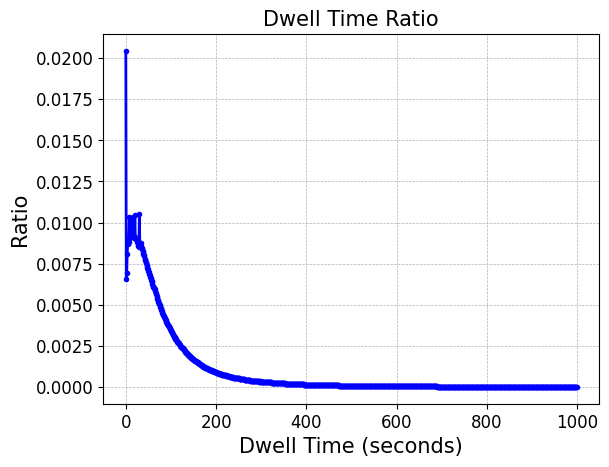} 
    }
    \subfigure[Dwell Time Ratio Log Line Graph]{
        \includegraphics[width=0.23\linewidth]{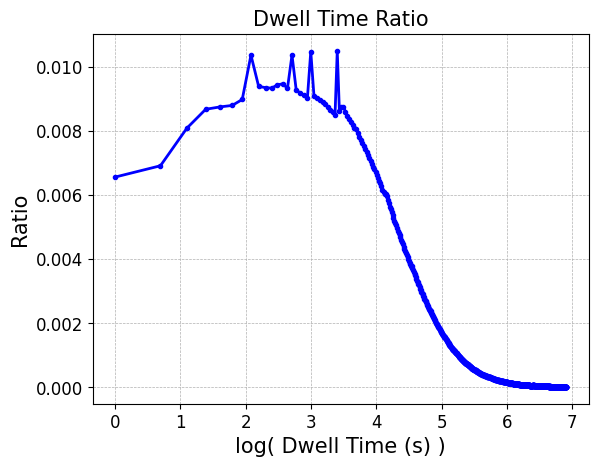} 
    }
    \subfigure[Dwell Time Ratio Pie Graph]{
        \includegraphics[width=0.23\linewidth]{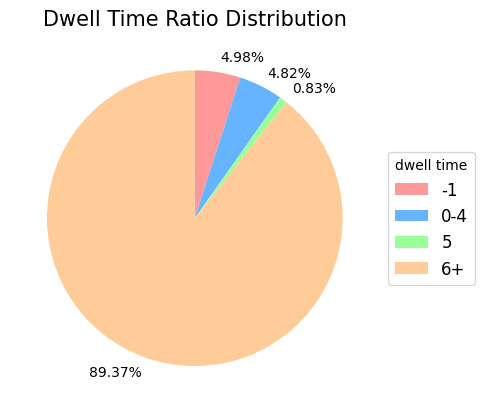}
    }
    \subfigure[Dwell Time Ratio Bar Graph]{
        \includegraphics[width=0.23\linewidth]{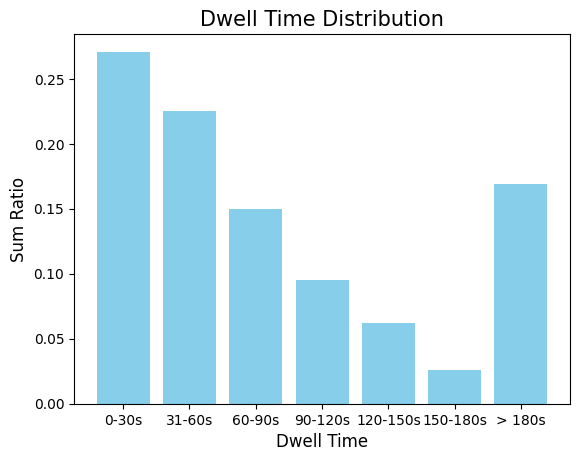}
    }
    \caption{ Dwell time distribution.}
  \label{f2}

\end{figure*}

In this subsection, we focus on analyzing the distribution of user dwell time to underscore its significance. Initially, we selected a complete day's worth of user log data from the MSN News website on July 18, 2023, recording the dwell time for each user click. Due to the nature of online data collection, some behaviors were not captured and are therefore denoted as \textbf{-1}, while the rest of the dwell times were subjected to a floor operation. 

As shown in Fig. \ref{f2}(a), the dwell time distribution is long-tailed, with the largest proportion being news items with a dwell time of zero. This observation confirms our earlier hypothesis that there is a substantial amount of click noise, suggesting that users' clicks may not align with their preferences and that relying solely on all click behaviors could potentially lead to suboptimal recommendation performance. 

Moreover, user clicks are predominantly concentrated between \textbf{5s} to \textbf{200s}, with excessively long dwell times possibly reflecting unintentional stays on a page without substantive reading. Additionally, as depicted in Fig. \ref{f2}(c), cases where dwell time is (-1) account for \textbf{4.97\%} of the data, indicating that nearly 1/20th of the user news click dwell times experience a collection delay. Therefore, during the model design process, the robustness of the model to dwell time must be considered to ensure accurate predictions despite potential delays. It is worth mentioning that \textbf{89.37\%} of user news clicks exceed \textbf{5s}, providing a solid data foundation for defining "effective clicks" in subsequent discussions. 

Furthermore, as depicted in Fig. \ref{f2}(d), we divided the data into segments with 30-second intervals, and as time increases, the proportion of user dwell time correspondingly decreases. The 0-30 second range constitutes a substantial share, reaching nearly one-quarter of the total, whereas the 150-180 second range drops to approximately 2.5\%. This indicates that people's reading habits tend to resemble browsing behavior, with a significant amount of weak feedback clustered within specific tie spans. Consequently, this presents a challenge for models to effectively differentiate between positive and negative user preferences.

In summary, user dwell time can significantly differentiate between types of user click behaviors, eliminate click noise, and enhance the accuracy of user modeling. However, when deploying in real-world recommendation scenarios, the model's robustness to dwell time is crucial to maintain strong recommendation performance, even when there are delays in dwell time collection.

\subsection{Problem Definition}
For the definition of pre-trained news recommendation pipline, given a user's historical behaviors, and candidate news information, the our model first obtain a candidate news understanding representation ($\mathbf{N}$) by Pre-trained Language Models and extraction the user interests representation ($\mathbf{U}$) from their user's historical clicked behaviors ($\mathbf{B}_u$), with clicked news representation ($\mathbf{N}_i$) by Pre-trained Language Models. The final goal is to predict the probability of whether the user will click on the candidate news that they have not seen before.

As previously shown, user dwell time is vital for improving the precision of user interest modeling. In this study, we integrate user dwell time data into the news comprehension and user interest modeling components of conventional models. This integration is designed to address the limitations associated with insufficient feedback from click interactions. 

\section{Methods}  
\begin{figure}[t]
  \includegraphics[width= \linewidth]{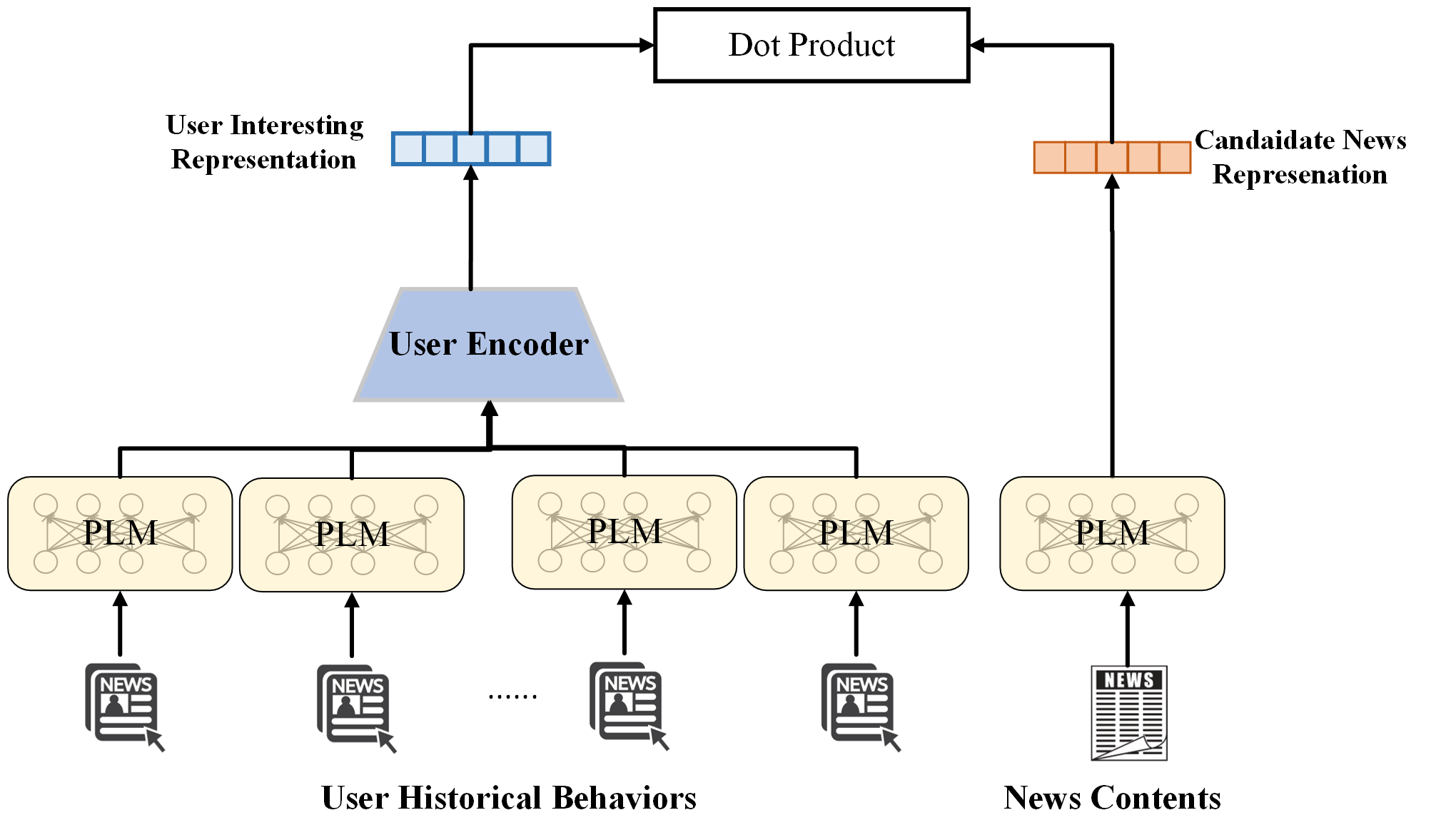}
  \caption{\textbf{The structure of PLM-NR.}}
  \label{f3}
\end{figure}

In this section, we outline our strategy for incorporating dwell time information to enhance the efficacy of the news recommendation model on the MSN News website\footnote{https://www.msn.com/en-us/news}. We start by describing our cutting-edge baseline model as it is applied in an industrial setting. Following this, we propose two novel approaches for the practical application of dwell time integration. These approaches are crafted to optimally utilize dwell time data, thus improving both the precision and relevance of news recommendations on the MSN platform.

\subsection{Model Structure}
In this subsection, we divided model into news understanding and user modeling. 

\textbf{News Understanding.} As large language models (LLMs) have demonstrated impressive capabilities in natural language processing, their strong power in understanding text is under using. Meanwhile, the owners of these LLMs have started offering Embedding as a Service (EaaS) to assist model with downstream NLP tasks. Therefore, we exploit the Text-Embedding-Ada-002 from OpenAI to get our news understanding embedding ($\mathbf{N} \in \mathcal{R}^d $). 

\textbf{User Modeling.}  In real-world recommendation scenarios, we propose two methods to extract user interests from historical click behaviors. Initially, we utilize the OpenAI API to acquire embeddings of news items clicked by the user and construct the user matrix \(\mathrm{E}_{u} = \mathbf{Stack} [\mathrm{N}_{1},\mathrm{N}_{2},... ,\mathrm{N}_{H}]\), where \(H\) represents the number of news items clicked by the user. Subsequently, drawing inspiration from \emph{NRMS}\cite{wu2019neural}, we have designed two distinct mechanisms: \emph{content attention} (as depicted in eq. \ref{eq1}) and \emph{multi-head attention}\cite{vaswani2017attention} (as illustrated in eq. \ref{eq2}). These mechanisms are employed to capture the interactions among the user's click behaviors, thereby enhancing the model's ability to understand and represent user interests more accurately and effectively.
\begin{equation}
\label{eq1}
\mathbf{U}_{context} = \mathbf{Softmax} (\mathrm{E}^{T}_{u} \times \mathrm{E}_{u}),
\end{equation}
\begin{equation}
\label{eq2}
\mathbf{U}_{context} = \mathbf{MultiHead} (\mathrm{E}_{u}, \mathrm{E}_{u}, \mathrm{E}_{u}),
\end{equation}
where $\mathrm{E}_{u} \in \mathcal{R}^{H \times D}$ is user clicked matrix and  $\mathrm{U}_{context} \in \mathcal{R}^{H \times D}$ is user context matrix. 
Final, we place a attention pooling network (can be simplied as $ \mathbf{AttPool(\cdot)} $ in the following paper) to get the user interest representation with Eq. \ref{ep3}. 
\begin{equation}
\begin{gathered}
\label{ep3}
   \alpha^{n} = \mathbf{Softmax} \left( \mathbf{Tanh} \left(\mathbf{W} \times \mathbf{E}^{i}_{u} + \mathbf{b} \right) \right), \
   \mathbf{U}=\sum_{i=1}^{N} \alpha_{i}^{n} \mathbf{N}_{i}^{n}
\end{gathered}
\end{equation}
where $\alpha^{n} \in \mathcal{R}^{D}$ is attention value of each user clicked news, $\mathbf{W} \in \mathcal{R}^{D \times D}$, $ \mathbf{b} \in \mathcal{R}^{D}$ are trainable parameters in MLP layer.

\textbf{Prediction.} 
The predicted likelihood is calculated by taking the dot product of user and news representations, 
\begin{equation}
\label{eq4}
\hat{y}_{ui} = {\mathbf{N}_{c}} \mathbf{U}^\mathrm{T},
\end{equation}
where $\mathbf{N}_{c}$ denotes the representations of the candidate news, $\mathbf{U}$ denotes the user interest representations, and $\mathrm{T}$ indicates the vector transpose operation.

\subsection{Dwell Time Enhanced User Modeling}
\begin{figure*}[tp]
  \includegraphics[width= \linewidth]{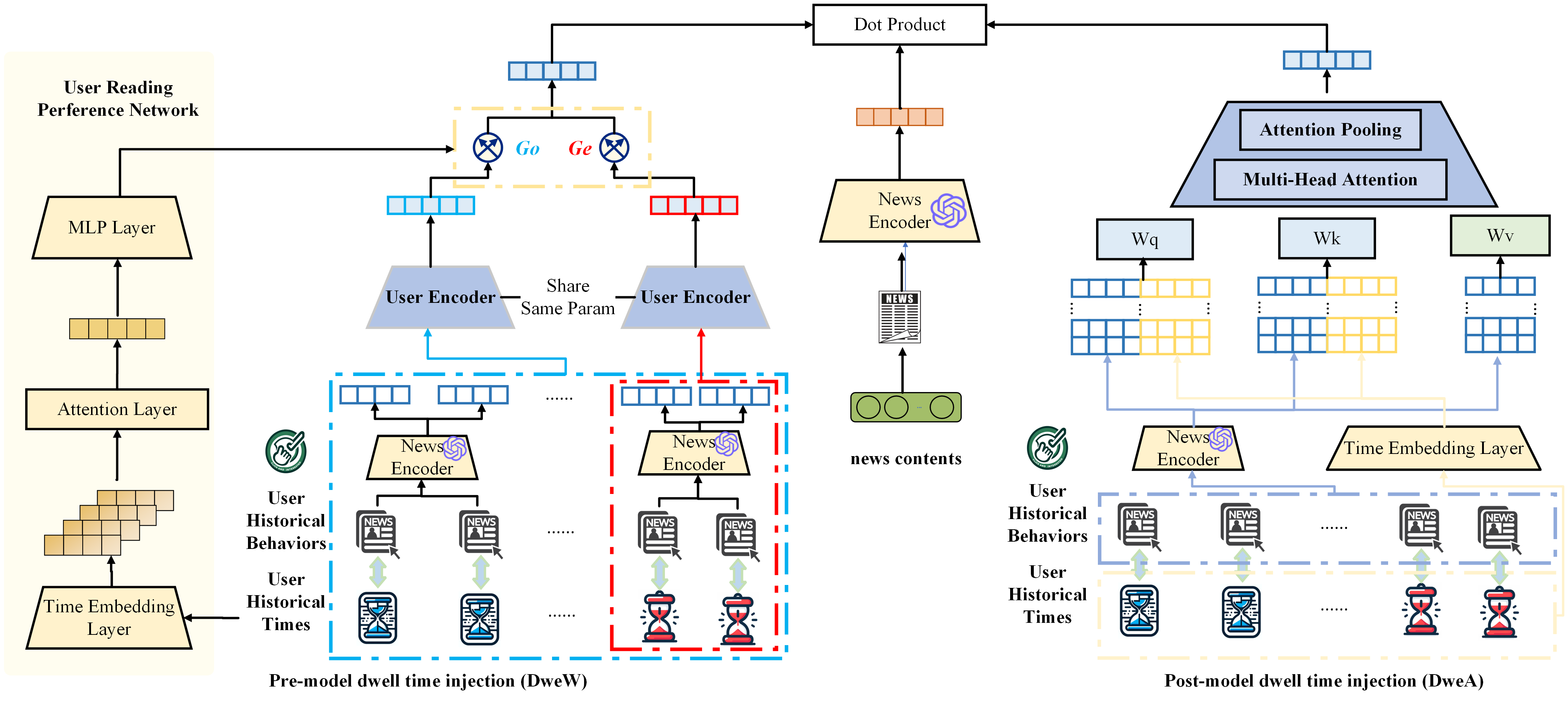}
  \caption{\textbf{The overall structure of our methods.}}
  \label{f4}
\end{figure*}

Building on our baseline architecture, we aim to leverage dwell time data to improve model performance. We introduce two strategies for incorporating dwell time: pre-model dwell time injection (\textbf{DweW}) and post-model dwell time injection (\textbf{DweA}). Additionally, we emphasize the robustness of our model in handling dwell time information, ensuring approximate predictions when user click dwell time data is unavailable. The model's structure is depicted in Fig. \ref{f4}, with \textbf{DweW} on the left and \textbf{DweA} on the right.

\subsubsection{Pre-injection of dwell time}
The NRNF model uses dwell time as a threshold for feedback polarity, but equating short dwell times solely with negative feedback is insufficient. This approach neglects individual user habits and doesn't reliably link shorter engagement to content dislike. It also exhibits a vulnerability to dwell time variations, undermining its stability.

In contrast, our method prioritizes the concept of \textbf{effective clicks} initiated by users and refine the foundational model of user interest with them. By doing so, we not only maintain system robustness but also significantly elevate the relevance of \textbf{effective clicks} actions in the context of user engagement. Therefore, we regard the split user behaviors into two sequences, $\mathbf{C}_u^o$ and $\mathbf{C}_u^e$, with Eq. \ref{eq4}. We follow the baseline process to construct user-news matrix $\mathbf{E}_u^o$ and $\mathbf{E}_u^e$ with OpenAI API.
\begin{equation} 
    \label{eq4}
    \mathbf{C}_u = 
    \begin{cases}
    \mathbf{C}_u^e & \mathrm{dwell_{n} > 5}, \\
    \mathbf{C}_u^o &  \mathrm{dwell_{n} > 0}
    \end{cases}    
\end{equation}
We then input $\mathbf{E}_u^o$ and $\mathbf{E}_u^e$ into same multi-head attention layer to get context information, respectively.
\begin{equation}
\begin{gathered}
\label{eq5}
\mathbf{U}_{context}^o = \mathbf{MultiHead} (\mathbf{E}_u^o, \mathbf{E}_u^o, \mathbf{E}_u^o), \
\mathbf{U}^o = \mathbf{AttPool} (\mathbf{U}_{context}^o), 
\end{gathered}
\end{equation}
\begin{equation}
\label{eq6}
\mathbf{U}_{context}^e = \mathbf{MultiHead} (\mathbf{E}_u^e, \mathbf{E}_u^e, \mathbf{E}_u^e),\
\mathbf{U}^e = \mathbf{AtnPool} (\mathbf{U}_{context}^e), 
\end{equation}

Considering that each user has unique personalized reading habits, we have designed a User Reading Preference Network to analyze user reading time behavior. Subsequently, we can determine the significance of \textbf{effective click} actions for a specific user and adjust their interest representation based on the original one. We first embed user dwell time in to matrix ($ \mathbf{D}_u $) and exploit attention layer to obtain the weight ($ \mathbf{G} \in \mathcal{R} ^ {2} $) for origin and effective representation.
\begin{equation}
\begin{gathered} 
\label{ep7}
\mathbf{G} = \mathbf{Softmax} \left( \mathbf{Tanh} \left(\mathbf{W}_d \times \mathbf{D}_u + \mathbf{b}_d \right) \right), \\
\mathbf{U} = \mathbf{G}_e \times \mathbf{U}^e + \mathbf{G}_o \times \mathbf{U}^o,
\end{gathered}
\end{equation}

\subsubsection{Post-injection of dwell time}
The CPRS model is a solution that incorporates dwell time information into user modeling to refine attention values. However, we contend that the model is overly dependent on dwell time data, lacking robustness as a result. Furthermore, the model treats semantic-level and dwell-time-level attention separately, failing to consider their interaction. This oversight is significant, as semantics and timing frequently exhibit a strong correlation in real-world scenarios. 

Therefore, in our \textbf{DweA} model, we aim to let user modeling network to \textbf{Aware} the dwell time information and exploit the strong power to capture the strong relationship between semantic and timing. It is a simple but insight and effective solution. Specifically, we obtain a user dwell time embedding matrix ($\mathbf{D}_u$) through a dwell time embedding layer, as same with \textbf{DweW}. Subsequently, we concatenate this matrix with the original user-news matrix $\mathrm{E}_{u}^a = \mathbf{Stack} [\mathrm{E}_{u},\mathbf{D}_u] $, where ${E}_{u} \in \mathcal{R}^{H \times (d + d_{dw})}$
is serve as the input for the $ \mathbf{Q} $ and $ \mathbf{K} $ matrices in the \emph{multi-head attention network} to revise attention value in user behaviors, as shown in Eq. \ref{eq8}. 
\begin{equation}
    \label{eq8}
 \begin{gathered}    
 \mathbf{U}_{context}^a = \mathbf{MultiHead} (\mathbf{Stack} [\mathrm{E}_{u},\mathbf{D}_u], \mathbf{Stack} [\mathrm{E}_{u},\mathbf{D}_u], \mathbf{E}_u), \\ \mathbf{U}^a = \mathbf{AttPool} (\mathbf{U}_{context}^a), 
    \end{gathered}
\end{equation}

It is important to note that we exclusively modify the inputs for the $ \mathbf{Q} $ and $ \mathbf{K} $ matrices, as these dictate the relevance of other users' click behaviors to a particular action, mirroring the core function of dwell time data. Moreover, in scenarios where dwell time information is being collected, our model maintains its efficacy through semantic-level representation.

\subsection{Model Training}
We use a \textbf{negative sampling} technique to balance positive and negative items. For each user clicked news, we considered it a positive sample. Besides, we randomly sample $\mathbf{K}$ news presented in the same session but are not clicked by this user as negative samples. Furthermore, we disrupt the order of news to avoid possible positional bias. The click probability scores of positive and negative news are expressed as
\begin{math}
\mathbf{{y}^{+}} 
\end{math}
and
$[\hat{{y}_{1}^{-}}, \hat{{y}_{2}^{-}}, ..., \hat{{y}_{k}^{-}}]$. These scores are normalized by the $\mathbf{Softmax}$ function. Therefore, we convert the problem of predicting the clicking likelihood into a perudo $\textbf{k+1}$ classification task. The loss function is shown in the Eq. \ref{31}.
\begin{equation}
\label{31}
\mathcal{L}=-\sum_{i \in \mathcal{S}} \log \left(\frac{\exp \left(\hat{y}_{i}^{+}\right)}{\exp \left(\hat{y}_{i}^{+}\right)+\sum_{j=1}^{K} \exp \left(\hat{y}_{i, j}^{-}\right)}\right)
\end{equation}

\section{Experiment}
 The experiments are designed to answer the following questions:
\begin{itemize}[leftmargin= 0 pt,topsep =0 pt]
    \item \textbf{RQ 1}: Does our solutions have better recommendation results than the outstanding PLM-NRline models?
    \item \textbf{RQ 2}: Does our solutions have better dwell time information robustness than the other dwell time-PLM-NRd models?
    \item \textbf{RQ 3}: How effective is the our models in predicting 'effecitve instances'?
\end{itemize}

\subsection{Experimental Settings}
\subsubsection{Datasets}
For fit the real-worlds recommendation situation in MSN news websites\footnote{\url{https://msnews.github.io/index.html}} \cite{wu2020mind}, we use a proprietary dataset collected by our self from a commercial MSN news platform (denoted as MSN News Dataset). It contains whole day logs of 53,497,936 users and 5,545,472 news. We use data in 13/07/2023 for training and 03/08/2023 for testing. More detailed statistical information of the two datasets is shown in Table \ref{datasetdetail}.

\begin{table}[h] 
  \caption{\textbf{Detailed statistics of real-world datasets.}}
  \label{datasetdetail}
  \begin{tabular}{cc|cc}
    \toprule 
    \multicolumn{4}{c}{MSN News Dataset} \\
    \midrule
    \textbf{$\#$Users} & 53,497,936 & \textbf{$\#$Avg. dwell time} & 105.3941 \\
    \textbf{$\#$News} & 5,545,472 & \textbf{$\#$Avg. title length} & 11.3753  \\ 
    \textbf{$\#$Impressions} &38,912,648,940 & \textbf{$\#$ Avg. body length} & 550.0322 \\  
    \textbf{$\#$Clicks} & 220,032,456 &  &  \\
    \bottomrule
    \end{tabular}
    \label{dataset}
\end{table}

\subsubsection{Parameter Settings}
During the model training phase, we employed the \textbf{Adam} optimizer~\cite{kingma2014adam} with a learning rate of ${1} \times {10}^{-3}$. The batch size was set to 32, and we applied a dropout rate of 0.2 to prevent overfitting. The negative sampling ratio $k$  was established at 4. To extract user interest representation, we selected the last 50 behaviors from user impressions. The multi-head attention structure was configured with 10 attention heads and an attention dimension of 20. For evaluating the model's performance, we chose metrics such as AUC, MRR, NDCG@5, and NDCG@10.

To facilitate efficient training, we utilized \textbf{8 V100 GPUs} over a 24-hour period and \textbf{1 V100 GPU} for tasks requiring approximately one hour. This setup ensured optimal resource allocation and maximized computational efficiency throughout the experiment.

As large language models (LLMs) demonstrate impressive capabilities in natural language pIMPessing, we use \textbf{Text-Embedding-Ada-002} \footnote{\url{https://platform.openai.com/docs/guides/embeddings}} to construct our news understanding representation in the news encoder, inputing the news title and abstract.

\subsection{Overall Performance}
In this section, we compare four excellent exsiting baseline model, PLM-NR and PLM-NRMS for industry baselines, NRNF and CPRS for dwell time injection baselines.
The comparison methods include:
\begin{itemize}[leftmargin= 12 pt,topsep =0 pt]

\item \textbf{NRNF}\cite{wu2020neural}: A news recommendation model that utilizes dwell time information to discern between positive and negative click behaviors.

\item \textbf{CPRS} \cite{cprs}: A news recommendation model that employs dwell time data to adjust the attention weights among user click behaviors.

\item \textbf{PLM-NR} \cite{plm-nr}: A pre-trained news recommendation model leveraging the OpenAI API to achieve powerful representations for both candidate and clicked news items.

\item \textbf{PLM-NRMS} \cite{wu2019}: A pre-trained news recommendation model that uses the OpenAI API in conjunction with a transformer structure to capture user interest representation.

\end{itemize}

\begin{table*}[t]
\caption{Overall experiments on two type of self-constructed datasets.}
\label{t2}

    \begin{tabular}{l|r|cccc|cccc}
    \toprule
    & & \multicolumn{4}{c}{Normal Dataset}  & \multicolumn{4}{c}{Real Dataset}\\
    & Model & AUC & MRR & NDCG@ & NDCG@10 & AUC & MRR & NDCG@ & NDCG@10\\
    \midrule
    \multirow{4}{*}{\textbf{Baselines}}
    & NRNF \cite{wu2020neural} & 63.53& 31.95 & 34.31 & 38.82 & 63.98 & 32.26 & 34.66 & 39.17 \\
    & CPRS \cite{cprs} & 66.99 & 34.48 & 36.98 & 41.38 & 67.29 & 34.70 & 37.23 & 41.63 \\
    & PLM-NR \cite{plm-nr} & 67.64 & \underline{35.12} & \underline{37.62} & \underline{41.99}  & 67.93 & \underline{35.34} & \underline{37.87} &  \underline{42.23} \\
    & PLM-NRMS \cite{wu2019} & \underline{67.79} & 35.06 &  37.60 & 41.91  & \underline{68.07} & 35.27 & 37.83 & 42.15 \\
    \hline
    \multirow{4}{*}{\textbf{Our Methods}}
    & \textbf{DweA}  & \textbf{67.87} & \textbf{35.20} & \textbf{37.90} & \textbf{42.10} & \textbf{68.29} & \textbf{35.37} & \textbf{37.93} & \textbf{42.27} \\
     & IMP & $\uparrow 0.44\%$ & $\uparrow 0.27\%$ & $\uparrow 0.53\%$ & $\uparrow 0.38\%$ & $\uparrow 0.44\%$ & $\uparrow 0.44\%$ & $\uparrow 0.66\%$ & $\uparrow 0.47\%$ \\
    & \textbf{DweW} & \textbf{68.09} & \textbf{35.22} & \textbf{37.82} & \textbf{42.15} & \textbf{68.37} & \textbf{35.50} & \textbf{38.12} & \textbf{42.43} \\
    & IMP & $\uparrow 0.12\%$ & $\uparrow 0.22\%$ & $\uparrow 0.74\%$ & $\uparrow 0.26\%$ & $\uparrow 0.32\%$ & $\uparrow 0.08\%$ & $\uparrow 0.16\%$ & $\uparrow 0.09\%$ \\
    \bottomrule
    \end{tabular} 
\end{table*}

\begin{table*}[htp]

  \caption{Robustness validation experiment with mask dwell time injection.}
  \label{t3}
   \begin{tabular}{l|r|cccc|cccc}
    \toprule
    & & \multicolumn{4}{c}{Normal Dataset}  & \multicolumn{4}{c}{Real Dataset}\\
  \midrule
     & Model & AUC & MRR & NDCG@5 & NDCG@10 & AUC & MRR & NDCG@5 & NDCG@10 \\
  \midrule
  & \textbf{Best PLM-based NR} & 67.79 & 35.12 &37.62 & 41.99 & 68.07  & 35.34 & 37.87 & 42.23\\
  \midrule
  \multirow{4}{*}{\textbf{Baselines}} 
   & \textbf{NRNF} -w/o di &  63.83 & 32.10 &  34.46 &	38.93  & 64.18 & 32.35 & 34.72 &	39.21\\
   & GTB & \textbf{-0.0584} & \textbf{-0.0860} & \textbf{-0.0840} &	\textbf{-0.0729} & \textbf{-0.0571} & \textbf{-0.0846} & \textbf{-0.0832} & \textbf{-0.0715} \\
   &\textbf{CPRS} -w/o di & 67.00 & 34.44 & 36.95 & 41.29	 & 67.32 & 34.71 & 37.24 &	41.59 \\
   & GTB & \textbf{-0.0117}  & \textbf{-0.0194} & \textbf{-0.0178} & \textbf{-0.0167} & \textbf{-0.0110} & \textbf{-0.0178} & \textbf{-0.0166} & \textbf{-0.0152}	 \\
  \midrule
   \multirow{4}{*}{\textbf{Our Methods}} 
   & \textbf{DweA} -w/o di & 67.71 & 34.91 & 37.44 & 41.79 & 68.00 & 35.14 & 37.70 & 42.04 \\
   & GTB &  \textbf{-0.0012} &  \textbf{-0.0060} &  \textbf{-0.0048} &	\textbf{-0.0048} &  \textbf{-0.001} &  \textbf{-0.0057} & \textbf{-0.0045} & \textbf{-0.0045} \\
   & \textbf{DweW} -w/o di & 67.54  & 34.80 & 37.30 & 41.67 &  67.78 & 34.99 & 37.51 & 41.89 \\
   & GTB & \textbf{-0.0037} & \textbf{-0.0091} & \textbf{-0.0085} & \textbf{-0.0076} & \textbf{-0.0043} & \textbf{-0.0099} & \textbf{-0.0095} & \textbf{-0.0081}	 \\
  \bottomrule
  \end{tabular}
\end{table*}

\subsubsection{Dataset Construction}
In this experiment, we construct two datasets to evaluate our model. The \textbf{Normal} dataset consists of test instances that do not consider dwell time, while the \textbf{Real} dataset is filtered to include only effective clicks with dwell times exceeding 5 seconds.
The rationale behind employing two test datasets is two-folds: Firstly, we strive for our model to excel in making recommendations amidst the noisy, real-world scenarios. Secondly, we aim to identify and recommend high-quality, highly engaging news articles that captivate users within the MSN platform product.

\subsubsection{Expeirment results}
The experimental results are presented in Table \ref{t2}, where \textbf{IMP} denotes the degree of improvement of our model compared to the baseline model, with an upward arrow indicating enhancement. As observed, taking the AUC metric as an example, DweA and DweD achieve improvements of \textbf{0.44\%} and \textbf{0.12\%} on the Normal dataset, respectively, and enhancements of \textbf{0.44\%} and \textbf{0.32\%} on the Real dataset, respectively.

Therefore, we have following conclusion according to the overall experiment results.
\begin{itemize}[leftmargin= 0 pt,topsep =0 pt]
\item Pre-trained news recommendation models exhibit superior performance compared to traditional methods based on user dwell time. This enhanced performance is attributed to the pre-trained models' ability to capture more accurate and comprehensive representations of news understanding. They can distinguish between genuine user interest and click noise at a higher-dimensional level, leading to a more precise depiction of user interests. Additionally, the notable performance of the PLM-NRMS model stems from its ability to capture the interactions between user behaviors, thereby helping to filter out low-quality or uninteresting news and accurately capture user preferences.
\item Dwell time injection-based news recommendation models did not exhibit commendable results in experiments. Firstly, the NRNF model requires the segmentation of the original user sequence into positive and negative sequences based on user dwell time, with the time threshold varying according to the specific data distribution. Finding the optimal time threshold demands considerable computational resources, often on a daily scale. Moreover, using dwell time as an absolute value to segment sequences fails to consider individual reading habits. Our method (DweW), through the User Reading Preference Network, allows the model to personalize weights for different users.
\item Additionally, the CPRS model did not yield favorable experimental outcomes. We believe its shortcoming lies in the physical separation of user dwell time and news semantics, overlooking their interaction. In reality, user dwell time is often strongly correlated with news content. For instance, factual news typically has shorter texts and, consequently, shorter dwell times but attracts widespread attention. Our method (DweA) introduces dwell time into the original news semantic interactions, taking into account their interplay and thereby achieving a more distinguished performance.
\item Ultimately, although pre-trained news recommendation models nearly match our proposed model's performance by capturing powerful news representations, the gap widens on real-world datasets. Our method focuses more on users' real interests, boosted by dwell time, leading to higher-quality news recommendations that genuinely resonate, rather than just capturing attention.
\end{itemize}

\subsection{Robustness validation experiments}
In our ablation study, we employed two approaches to demonstrate the robustness of our model to users' dwell time. Firstly, we incorporated click samples with unknown dwell times into a real dataset, proving that the model has enhanced predictive accuracy for such samples. Secondly, during testing, even when dwell time information was masked, the model still managed to make reasonably accurate predictions.

\subsubsection{For Masking Dwell Time Information Injection}

This experiment is designed to replicate real-world scenarios, focusing on the challenge of delayed user dwell time data collection. We test our model in an extreme case where no dwell time information is available, assessing its capacity to still make high-quality predictions.

The results is shown in Table \ref{t3}, outlined in the table and quantified by the "Gap to Baseline" (\textbf{GTB}) metric, demonstrate a minimal performance gap between our robust model and top-performing baseline models. Notably, the largest decrease in DweW's AUC is a mere \textbf{-0.0037} and \textbf{-0.0043}. This highlights the importance of dwell time in accurate user interest modeling, while showcasing our model's resilience in the absence of such data.

Our method shows greater robustness to dwell time variations compared to other models incorporating this data. It maintains similar performance to the PLM baseline model, indicating an ability to balance news semantics effectively, even without dwell time information. This adaptability positions our model as a strong contender for real-world recommendation scenarios, capably managing the complexities of variable user engagement data.

\subsubsection{For Predicting Samples with Unknown Dwell Time}
This section evaluates our model's efficacy in recommending samples with uncertain dwell times, simulating real-world scenarios. Integrating dwell time data may theoretically introduce a bias towards preferences for longer engagements. To address this, our model incorporates a perception-weighted mechanism to enhance robustness. We utilize a dataset comprising samples with undefined dwell times to construct a robustness-centric dataset.

\begin{table}[h]
    \caption{Robustness validation experiment with unknown dwell time instances.}
    \label{t4}
    \begin{tabular}{r|cccc}
    \toprule
     \multicolumn{4}{c}{\textbf{Robust} Dataset}\\
    Model & AUC & MRR & NDCG@ & NDCG@10\\
    \midrule
    NRNF \cite{wu2020neural} \cite{mai2024knowledge} & 63.84 &  31.96 & 34.29 & 38.79   \\
    CPRS \cite{cprs} & 67.27 & 34.58 &  37.04 & 41.44 \\
    PLM-NR \cite{plm-nr} & 67.84 &  \underline{35.13} &  \underline{37.65}  &  \underline{42.00} \\
    PLM-NRMS \cite{wu2019} & \underline{67.99} & 35.10 & 37.63 & 41.94 \\
    \hline
    \textbf{DweA} & 68.15 & 35.16 & 37.68 & 42.01 \\
    IMP & $\uparrow 0.24\%$ & $\uparrow 0.09\%$ & $\uparrow 0.08\%$ & $\uparrow 0.02\%$ \\
    \textbf{DweW} & 68.21 & 35.19 & 37.71 & 42.06 \\
    IMP & $\uparrow 0.32\%$ & $\uparrow 0.17\%$ & $\uparrow 0.16\%$ & $\uparrow 0.14\%$ \\

    \bottomrule
    \end{tabular} 
\end{table}

Table \ref{t4} highlights the impact of incorporating click samples with unknown dwell times on model performance, demonstrating a decrease across all models and emphasizing the challenge of accurately predicting such samples. Notably, our DweA and DweW models exhibit remarkable resilience, maintaining superior recommendation quality despite this challenge. Analysis reveals that DweA and DweW outperform other models in uncertain dwell time conditions, achieving impressive improvements of \textbf{0.24\%} and \textbf{0.32\%} in AUC over the leading baseline models. This noteworthy performance underscores their ability to deliver precise recommendations regardless of dwell time variations. These results affirm the effectiveness of our methods in ensuring high-quality news recommendations and their adaptability in discerning news semantics, essential for effective task management across diverse contexts.

\begin{figure*}[tp]
  \includegraphics[width= \linewidth]{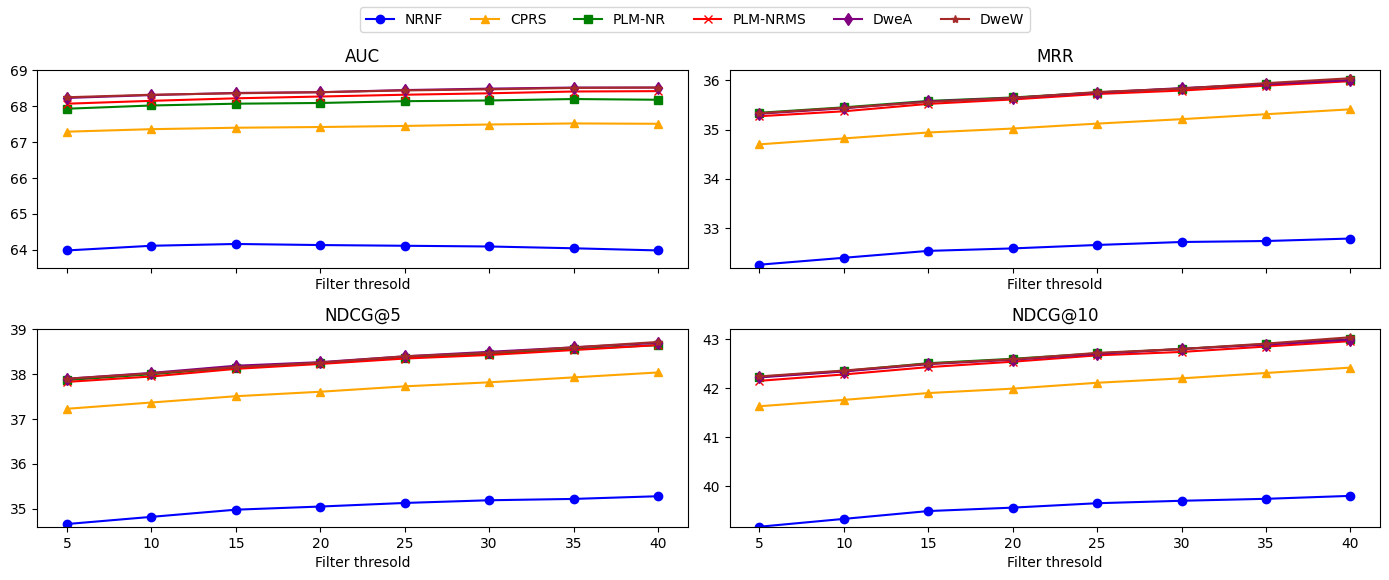}
  \caption{\textbf{Performance Comparison of Recommendation Algorithms Across Different Filter Thresholds.}}
  \label{fg5}
\end{figure*}

\subsection{Parameter sensitivity experiments}
In this experimental phase, we meticulously fine-tuned the hyperparameters within the \textbf{Real} dataset. This included adjusting the high-quality click filtering threshold in 5-second increments, ranging from 5 to 40 seconds. The objective was to assess the model's consistent performance over diverse news dwell time intervals.

The results, depicted in Fig. \ref{fg5} and measured by AUC, MRR, NDCG@5, and NDCG@10, demonstrate a significant enhancement across all model metrics as dwell time thresholds increase, with the exception of NRNF's AUC. This observation suggests that the models adeptly match high-quality news items with user preferences, leading to more precise recommendations compared to lower-quality news sources. Moreover, it highlights the potential for clicks, often construed as weak feedback, to occasionally lead models astray, thereby impacting the faithful representation of user preferences.
 
Furthermore, \textbf{DweA} and \textbf{DweW} consistently outperform the baseline models across all recommendation metrics, irrespective of the threshold criteria, notably surpassing the current state-of-the-art PLM-NRMS in MSN platform. Specifically, in terms of AUC, they exhibit an average improvement of approximately \textbf{0.1}. This performance underscores not only the robustness of our model in handling dwell time information but also its superiority in recommending news samples with diverse dwell times, thus affirming the model's effectiveness.

\section{Conclusions}
This paper initially highlights the shortcomings of current high-performing pre-trained news recommendation models, specifically the noise and bias present in modeling user interests through click behaviors. Existing high-performance network structures are still unable to precisely eliminate these inaccuracies. Therefore, we propose the use of user dwell time information to assist in filtering high-quality news that aligns with user interests, thereby capturing authentic user preferences. Moreover, considering the existing time infusion methods' lack of robustness to time data, we inject time information through self-awareness within the model, ensuring its resilience and suitability for real-world recommendation scenarios with potential time information delays. Specifically, we introduce two simple yet effective strategies for dwell time information injection, named \textbf{DweA} and \textbf{DweW}, which infuse time information before and after input, prompting the model to adjust attention values and network weights based on the user dwell time information sequence. Through extensive experimentation, our methods are proven to be effective not only in standard datasets but also in real datasets, i.e., datasets devoid of click noise, where our approach significantly outperforms existing SOTA pre-trained news recommendation models. More importantly, our model maintains high robustness even in the absence of user dwell time information, making it suitable for the complex recommendation scenarios encountered in the industrial sector.


\bibliographystyle{acm}
\bibliography{ref.bib}
\end{document}